\documentclass[sigplan,screen]{acmart}

\copyrightyear{2025}
\acmYear{2025}
\setcopyright{acmlicensed}\acmConference[ASPLOS '25]{Proceedings of the 30th ACM
International Conference on Architectural Support for Programming Languages and
Operating Systems, Volume 1}{March 30-April 3, 2025}{Rotterdam, Netherlands}
\acmBooktitle{Proceedings of the 30th ACM International Conference on Architectural
Support for Programming Languages and Operating Systems, Volume 1 (ASPLOS '25),
March 30-April 3, 2025, Rotterdam, Netherlands}
\acmDOI{10.1145/3669940.3707266}
\acmISBN{979-8-4007-0698-1/25/03}

\usepackage{afterpage}
    
\usepackage[ruled,linesnumbered,noend]{algorithm2e}
\usepackage{amsmath,amsfonts,amsthm}
\usepackage{array}
\usepackage{booktabs}
\usepackage{caption}
\usepackage{color}
\usepackage[inline]{enumitem}
\usepackage{endnotes}
\usepackage{float}
\usepackage{graphicx}
\usepackage[utf8]{inputenc}
\usepackage{makecell}
\usepackage{multirow}
\usepackage{natbib}
\usepackage{siunitx}
\usepackage{subcaption}
\usepackage{fancyvrb}
\usepackage{xspace}
\usepackage{listings}
\usepackage{xcolor}

\newcommand{\sys}{{S\small{ING}}\xspace}

\usepackage{tikz}
\newcommand*\circled[1]{\tikz[baseline=(char.base)]{
            \node[shape=circle,draw,inner sep=0.5pt] (char) {#1};}}
            
\DeclareMathAlphabet{\mathbfcal}{OMS}{cmsy}{b}{n}
\DeclareMathAlphabet{\mathcal}{OMS}{cmsy}{m}{n}
\renewcommand{\b}[1]{\textbf{#1}\xspace}
\renewcommand{\i}[1]{\textit{#1}\xspace}
\renewcommand{\t}[1]{\vspace{3pt}\noindent\textbf{#1}\xspace}
\newcommand{\f}[1]{\vspace{3pt}\noindent\underline{\textbf{#1}}\xspace}

\newenvironment{ul}{
\begin{itemize}[topsep=2.5pt, partopsep=0pt, leftmargin=1em]
  \setlength{\itemsep}{2.5pt}
  \setlength{\parskip}{0pt}
  \setlength{\parsep}{0pt}
}{\end{itemize}}

\newenvironment{ol}{
\begin{enumerate}[topsep=2.5pt, partopsep=0pt, leftmargin=1em]
  \setlength{\itemsep}{2.5pt}
  \setlength{\parskip}{0pt}
  \setlength{\parsep}{0pt}
}{\end{enumerate}}

\begin{document}
    \title{Design and Operation of Shared Machine Learning Clusters on Campus}
    
    \author{Kaiqiang Xu}
    \affiliation{%
      \institution{Hong Kong University of Science and Technology}
      \city{Hong Kong}
      \country{Hong Kong}
    }
    \email{kxuar@cse.ust.hk}

    \author{Decang Sun}
    \affiliation{%
      \institution{Hong Kong University of Science and Technology}
      \city{Hong Kong}
      \country{Hong Kong}
    }
    \email{dsunak@cse.ust.hk}

    \author{Hao Wang}
    \affiliation{%
      \institution{Hong Kong University of Science and Technology}
      \city{Hong Kong}
      \country{Hong Kong}
    }
    \email{hwangdv@connect.ust.hk}

    \author{Zhenghang Ren}
    \affiliation{%
      \institution{Hong Kong University of Science and Technology}
      \city{Hong Kong}
      \country{Hong Kong}
    }
    \email{zrenak@cse.ust.hk}

    \author{Xinchen Wan}
    \affiliation{%
      \institution{Hong Kong University of Science and Technology}
      \city{Hong Kong}
      \country{Hong Kong}
    }
    \email{xinchen.wan@connect.ust.hk}

    \author{Xudong Liao}
    \affiliation{%
      \institution{Hong Kong University of Science and Technology}
      \city{Hong Kong}
      \country{Hong Kong}
    }
    \email{xliaoaf@connect.ust.hk}

    \author{Zilong Wang}
    \affiliation{%
      \institution{Hong Kong University of Science and Technology}
      \city{Hong Kong}
      \country{Hong Kong}
    }
    \email{zwangfb@connect.ust.hk}

    \author{Junxue Zhang}
    \affiliation{%
      \institution{Hong Kong University of Science and Technology}
      \city{Hong Kong}
      \country{Hong Kong}
    }
    \email{jzhangcs@connect.ust.hk}

    \author{Kai Chen}
    \affiliation{%
      \institution{Hong Kong University of Science and Technology}
      \city{Hong Kong}
      \country{Hong Kong}
    }
    \email{kaichen@cse.ust.hk}

    \renewcommand{\shortauthors}{Kaiqiang Xu et al.}

    \begin{abstract}

    The rapid advancement of large machine learning (ML) models has driven universities worldwide to invest heavily in GPU clusters. Effectively sharing these resources among multiple users is essential for maximizing both utilization and accessibility. However, managing shared GPU clusters presents significant challenges, ranging from system configuration to fair resource allocation among users.

    This paper introduces \sys, a full-stack solution tailored to simplify shared GPU cluster management. Aimed at addressing the pressing need for efficient resource sharing with limited staffing, \sys enhances operational efficiency by reducing maintenance costs and optimizing resource utilization. We provide a comprehensive overview of its four extensible architectural layers, explore the features of each layer, and share insights from real-world deployment, including usage patterns and incident management strategies.
    
    As part of our commitment to advancing shared ML cluster management, we open-source \sys's resources to support the development and operation of similar systems.

\end{abstract}


    \maketitle
    \section{Introduction}
\label{section:introduction}

In the age of large machine learning (ML) models~\cite{gpt3, sd}, there has been a significant increase in the need for GPU clusters.
This surge in demand has been met with substantial research funding and human capital investments in universities throughout the world~\cite{nvidiasolution}.
However, not all campus infrastructures are equal, and many universities do not have the expertise to make the best use of shared GPU clusters. In most cases, campus IT (or graduate students) pick a tool that they are familiar with (e.g., Slurm~\cite{slurm}) and use it as-is with sub-optimal settings. Consequently, they end up underutilizing the already limited resources, putting a significant barrier to the democratization of access to ML. 

Effective management of a shared GPU cluster requires addressing a hierarchical set of system challenges, including cluster management frameworks~\cite{k8s, slurm}, scheduling strategies~\cite{pollux, sia}, network transport protocols~\cite{dctcp, mlt}, topology design~\cite{dcn, osa}, and other system configurations~\cite{conda, cuda}.
A brief online search yields numerous inquiries~\cite{reddit1, reddit2, loginnode3} for advice on managing shared GPU clusters.
Our observations indicate that fellow researchers invest significant effort in exploring software solutions, and frequently struggle with improving the clusters' \i{usability}, \i{stability}, and \i{performance} to meet expectations.

Admittedly, a wealth of information is available online offering recommendations~\cite{share1,share2} for managing shared clusters. However, effectively leveraging these solutions for a research-focused, shared GPU cluster demands deep expertise and continuous operational effort for evaluation, deployment, and maintenance.

Consider multi-user resource allocation as an example:
A simple and common approach to GPU sharing is to provide direct shell access to users, which can result in system instability arising from users concurrently running computing processes, competing for memory resources, potentially disrupting each other's runtime dependencies, and requiring manual coordination and careful cleanup.
On the flip side, Kubernetes~\cite{k8s}, a widely adopted cluster manager, is known to be effort-demanding in deployment. Originally designed for internet backend services, Kubernetes features a complex architecture with service-hosting capabilities like load balancing and rollbacks but lacks native support for user-access control and job queuing for resource sharing.
These complexities create engineering challenges and overhead when applied to ML model training jobs in campus ML clusters. 
Details of existing solutions are discussed in \S\ref{section:background}.

\vspace{5pt}

In this paper, we present our experience from the \i{design} and \i{operation} of \sys, a shared GPU cluster tailored for ML research in academic institutions.

\b{\i{Design}:}
\sys streamlines ML job processing into four system layers: job profile, adapter, scheduling, and execution~(\S\ref{section:design}). Each layer utilizes a low-maintenance software stack that efficiently handles tasks like job scheduling and user isolation, while minimizing burdens in operation. We employ effective resource allocation strategies to prevent low resource utilization and improve fairness during allocation.

\b{\i{Operation}:} 
Launched in early 2021, \sys has become a cornerstone of our institution's AI research infrastructure, managing over 160 GPUs and serving over 480 active users. These users have submitted over 28,000 ML jobs, which generated experiment results that contribute to about 40 peer-reviewed ML papers. 
We present a comprehensive analysis of \sys's usage patterns, job characteristics, and our incident handling experiences (\S\ref{section:usage}). We will also release job trace data to facilitate further research in this area~(\S\ref{section:opensource}).

At its core, \sys places a strong emphasis on simplicity and stability when it comes to selecting software stacks. In line with the constraints of limited manpower within research teams, our approach to designing and choosing \sys's components is guided by the principle of achieving \i{minimal viability} while still fulfilling the essential requirements for running ML jobs.
To facilitate this design goal, \sys adopts a 4-layer architectural abstraction, connecting its components that handle the compilation, scheduling, and execution of user-submitted jobs within the cluster. The detailed exploration of \sys's interface, internal workflow, and system configurations can be found in \S\ref{section:design}.

\sys offers a user-friendly interface from the local command line, allowing users to submit jobs with code, configurations, and specified resources. This workflow is also referred to as Machine-Learning-as-a-Service~(MLaaS)~\cite{mlaas}.
For maximum usability, \sys supports two job formats simultaneously: scripting (submitting a script, which is widely used with HPC clusters and adopted into ML computing) and containerizing (submitting a Dockerfile~\cite{dockerfile} describing a container that runs the job). 
Similar to cloud services providers, \sys automates resource provisioning for job execution, but with additional ML-specific features including dependency setup, job queuing and scheduling, port forwarding, interactive debugging, and access to logs and outputs.

\sys is designed with extensibility, allowing cluster operators to employ alternative options in the layers of \sys to achieve new capabilities. Beyond the default software selections that prioritize architectural simplicity and operational efficiency, we also discuss scaling opportunities such as expanding the executor backend to cloud platforms.

During developing and operating \sys, balancing user-friendliness with operational efficiency was challenging. We refined our design choices by monitoring user interactions and analyzing anonymous usage data, while addressing performance bottlenecks and resolving instances of failures and incidents. We delve into these operational insights, focusing on usage patterns, job characteristics, and incident handling and prevention in \S\ref{section:usage}.

\vspace{6pt}

This paper contributes to the field in two main aspects:

\begin{ul}
\item \b{Operation-Efficient Design} (\S\ref{section:design}): 
\sys stands out as an accessible solution that manages a shared GPU cluster and achieves operational efficiency (i.e., low maintenance cost and high resource utilization), specifically addressing the challenges posed by limited manpower in research teams. 
We describe seven core features of \sys's design that contribute to its operational efficiency.

\item \b{Operational Insights} (\S\ref{section:usage}): We track and analyze usage patterns and job characteristics and also share our experiences in incident handling during the operation of \sys.  These data and experiences offer valuable insights into the design and operation of similar campus ML clusters.
\end{ul}

\vspace{6pt}

This paper represents our ongoing effort aimed at improving the management of shared ML clusters in academic institutions. 
We make the source code, software stack configurations, and \sys job traces (\S\ref{section:usage:trace}) available to assist in deploying and operating similar facilities.

    \begin{table*}[ht]
    \footnotesize
    \centering
    
    \caption{Comparison of existing solutions for shared ML clusters. The analysis considers each solution’s core features without extensive customization. References to \#1-7 correspond to \sys's features described in \S\ref{section:design}.}
    \vspace{-6pt}

    \begin{tabular}{l|c|c|c|c}
    \hline
    \multicolumn{4}{l}{}\\[-8pt]
    \textbf{Feature Category} & \textbf{Direct Shell Access} & \textbf{Traditional HPC} & \textbf{Container Platforms} & \textbf{SING} \\ 
    \multicolumn{4}{l}{}\\[-8pt]
    \hline

    \multicolumn{4}{l}{}\\[-8pt]
    \multicolumn{4}{l}{\textbf{Resource Management}} \\ 
    \multicolumn{4}{l}{}\\[-9pt]
    \hline
    Job Resource Access & SSH \& Linux Permissions & Login Node Submission & Deploy Container via UI/CLI & User Local Submission (\#2)  \\ \hline
    Resource Scheduling  & Manual Coordination & FCFS, Queuing & Quota-based & Optimized FCFS (\#4)\\ \hline  
    
    \multicolumn{4}{l}{}\\[-8pt]
    \multicolumn{4}{l}{\textbf{Job Execution \& Monitoring}} \\ 
    \multicolumn{4}{l}{}\\[-9pt]
    \hline
    Work Description & Manual Command  & Job Script & Container File & \multirow{2}{*}{\makecell{Multi-Format Support (\#1) \& \\ Early-Init \& Caching (\#5)}}\\ \cline{1-4}
    Dependency Setup & Manual Command & Script Execution & Container Initialization & \\ \hline
    Job Monitoring & Server Login  & \multicolumn{2}{c|}{Offline Debug with Output Log } & Interactive Monitoring (\#3)\\ \hline 
    
    \multicolumn{4}{l}{}\\[-8pt]
    \multicolumn{4}{l}{\textbf{User Experience and Complexity}} \\ 
    \multicolumn{4}{l}{}\\[-9pt]
    \hline
    User Learning Curve & Simple - Basic Shell & Moderate - Job Script & Complex - Container Skills & Moderate - Intuitive Workflow (\#6) \\ \hline
    Operator Setup Cost & Low - Basic Linux  & Moderate - Install Software & High - Complex Architecture & \multirow{2}{*}{\makecell{Low to Moderate Cost (\S\ref{section:design:analysis}) \& \\ with Executor Extensibility (\#7)}}\\ \cline{1-4}
    Maintenance Cost & High - Stability / Conflicts & Moderate - Stability Issues & High - Limited ML Job Support  \\ \hline
    \end{tabular}
    
    \label{table:solutions}
\end{table*}

\section{Background and Motivations} 
\label{section:background}

The growing demand for GPUs has led to significant financial and human resource investments. 
However, effectively managing shared GPU clusters presents complex challenges, including resource and environment provisioning, usability and stability issues, and resource allocation efficiency.

\subsection{Existing Solutions for Shared ML Clusters}
\label{section:background:solutions}

In this section, we analyze the workflows of three common solutions for managing shared ML clusters. 
In Table~\ref{table:solutions}, we summarize their advantages and disadvantages.

\subsubsection{Direct Shell Access Sharing}

A common approach is to grant users native shell access to GPU-equipped machines through SSH, allowing direct interaction with GPU resources. 
However, due to the lack of resource isolation and management mechanisms, this method raises concerns about system stability and security. Without careful coordination, it can result in conflicts between competing user sessions, degrading system performance and making it difficult to ensure fair resource allocation.

Users typically follow the following workflow:

\i{Step 1}. User Connection: Users connect to GPU nodes via SSH, gaining direct access to GPU resources.

\i{Step 2}. Manual Environment Setup: Users are responsible for manually configuring their runtime environments, including installing necessary libraries and dependencies.

\i{Step 3}. Resource Allocation: Users must manually coordinate GPU resources through other channels, to avoid potential conflicts or performance degradation.

\i{Step 4}. Job Execution and Cleanup: Users initiate ML tasks by running their scripts directly on GPU nodes, requiring them to manually manage job execution and clean up the environment after the job finishes.

\t{Analysis.} While this approach offers simplicity in terms of user access, it presents significant drawbacks. 

First, users are required to manually configure their runtime environments, including the installation of necessary libraries and dependencies. This manual setup process can be error-prone and time-consuming, adding to the operational burden and potentially risking system stability.

Furthermore, users must proactively take responsibility for coordinating GPU sharing, initiating ML tasks, and cleaning up their environment afterward. It can potentially introduce conflicts between concurrent user sessions, which not only degrades system performance but also makes it challenging to ensure fair resource allocation.

\vspace{-3pt}
\subsubsection{Traditional HPC Solution}

Universities often adopt cluster workload management solutions from traditional High-Performance Computing (HPC) clusters, with the Slurm workload manager~\cite{slurm} being a popular choice.  
Slurm provides job queueing and scheduling capabilities in clusters.

In Slurm, users follow a more structured workflow:

\i{Step 1.} Job Submission: Users submit ML job requests from a login node to the Slurm, with resource request specified.

\i{Step 2.} Job Queueing: Slurm is responsible for job queuing and scheduling, using built-in schedulers.

\i{Step 3.} Runtime Environment: Users generally need to set up the runtime environments within the job script, as there is a lack of support for managing ML dependencies in Slurm.

\i{Step 4.} Execution: The system bootstraps and executes the ML job on allocated nodes, and releases resources afterwards.

\t{Analysis.} 
Slurm lacks support for provisioning the complex dependency environments required by ML jobs, posing engineering difficulties to individual users.
Job preparation and submission are required to be done on the login nodes of a Slurm cluster, and the login node often becomes a single point of failure~\cite{loginnode1,loginnode2}, due to unexpected user actions on the node (see detailed discussion in \S\ref{section:design:layer1}).
Furthermore, Slurm's coarse-grained resource allocation mechanisms may lead to resource fragmentations and underutilization under certain submission orders and job characteristics~\cite{fragmentation}.

\t{Opportunity.}
Originally tailored for MPI applications, Slurm stood out for its simplicity and performance. 
ML training jobs in research often resemble MPI applications, using advanced MPI techniques~\cite{nccl, horovod}. This similarity brings ML job scheduling within Slurm's scope. 
In this paper, our solution involves augmenting Slurm with ML-specific features that transform it into an effective tool for shared ML clusters.

\subsubsection{Container Platforms}
\label{section:background:container}

Container orchestration platforms such as Kubernetes~\cite{k8s} have garnered widespread adoption in the industry for managing large-scale distributed applications. 
Recently, Kubernetes has also been used to manage shared ML clusters, especially in production clusters with user-facing ML applications. 

In containerized solutions such as Kubernetes, users follow a workflow centered around container orchestration:

\i{Step 1.} Containerization: Users package their ML program and runtime dependencies into containers by writing a description file in a specific format.
    
\i{Step 2.} Container Deployment: Users submit container images on the Kubernetes cluster, specifying resource requests. 

\i{Step 3.} Job Scheduling: Kubernetes schedules pods on available nodes, managing resource allocation.
    
\i{Step 4.} Job Execution: Containers are initiated to execute ML tasks, with Kubernetes handling creation and releasing.

\t{Analysis.} 
Designed for internet services, Kubernetes has a complex architecture with features like service discovery, load balancing, and feature rollouts and rollbacks. In the context of ML research, these complexities are unnecessary and introduce engineering overhead.

More importantly, Kubernetes lacks native support for user-level access control and job queuing essential for shared ML clusters, each requiring a separate set of tools and configurations~\cite{k8s2} to be added on top of Kubernetes. Therefore, deploying and maintaining a Kubernetes setup demands considerable effort from cluster operators, including integration, testing, and ongoing maintenance. 

Meanwhile, for users \i{without prior experience}, creating containers can be error-prone~\cite{dockerfilesmell} with Dockerfile, and replicating a cluster environment locally for error reproduction is challenging and often requires operator assistance.

Overall, both users and operators face a steep learning curve with Kubernetes, requiring skill development to manage their inherent complexities effectively~\cite{k8s2}.

\t{Opportunity.} 
Container platforms like Kubernetes are powerful server operation tools, and \sys's design does not conflict with them. \sys uses an adapter layer that can operate over any execution layer, including container platforms or computing clouds, allowing resource extensibility while remaining transparent to the users.

\subsection{Cluster Schedulers and Computing Frameworks}

This section provides a brief overview of related systems that also operate in cluster environments. A more detailed discussion can be found in \S\ref{section:discussion}.

\t{ML Cluster Schedulers}. State-of-the-art schedulers, such as Sia~\cite{sia}, Pollux~\cite{pollux}, and Gavel~\cite{gavel}, optimize scheduling strategies to enhance cluster efficiency, such as reducing average job completion time. These schedulers can be integrated into cluster managers like Slurm, Kubernetes, and \sys, which manage the entire ML job lifecycle.

\t{Computing Frameworks}. Distributed computing frameworks such as Ray~\cite{ray} and Spark~\cite{spark} parallelize large jobs for efficient distributed execution. These distributed systems are programming frameworks used directly in user code, which can be submitted as a part of the user's job to cluster managers. These systems also support an optional "cluster mode" similar to a cluster manager and we discuss this in \S\ref{section:discussion}.

\subsection{Motivations and Requirements}

Current ML cluster management solutions reveal limitations in usability and operational efficiency in multi-tenant environments.
\sys's design is motivated by the need for a simple, stable, and efficient solution for managing shared ML clusters in research institutions.

An effective solution should address three objectives:

\begin{ol}
    \item \textbf{Reduce Operational Complexity}: Leverage a stable and low-maintenance architecture to reduce operational workload for cluster operators compared to existing solutions.
    \item \textbf{Improve User Accessibility}: Provide a clear and accessible interface that simplifies ML job submission and monitoring, reducing the learning curve for users.
    \item \textbf{Enable Fair and Efficient Resource Allocation}: Ensure fair resource allocation to maximize utilization while maintaining system responsiveness. 
\end{ol}

\sys addresses these requirements with a full-stack architecture designed for ML workflows, introducing features such as multi-format job support, optimized first-come-first-serve (FCFS) scheduling, and environment caching.  
\sys aims to create a more stable and user-friendly environment for shared ML cluster operations.

    \begin{figure}[t]
    \centering
    \includegraphics[width=\columnwidth]{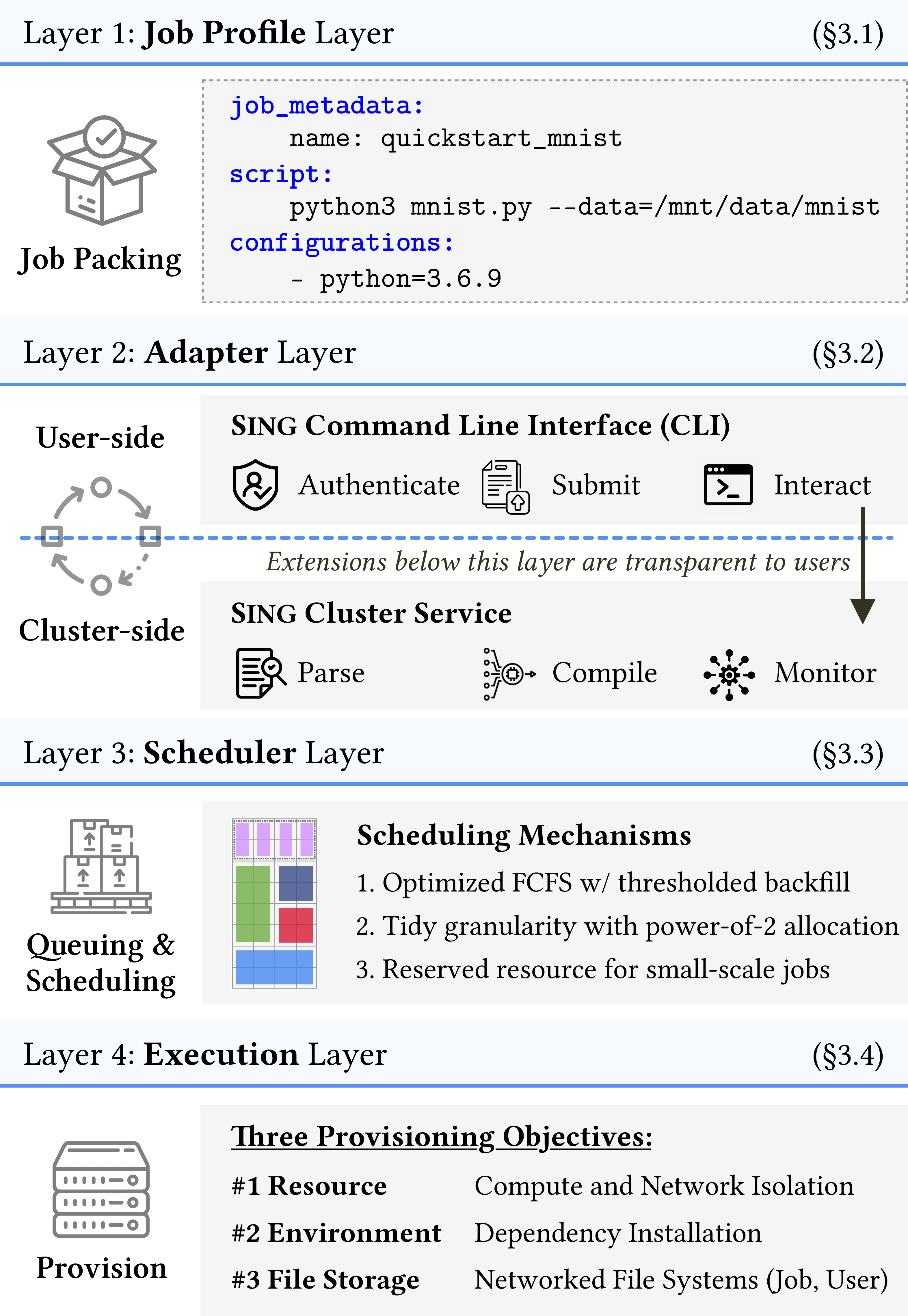}
    \caption{\sys uses 4-layer architecture to enhance the overall efficiency and flexibility in operation. }
    \label{figure:design}
\end{figure}

\section{Design and Implementation}
\label{section:design}

\sys's core architecture design follows a four-layer abstraction to define, construct, and execute user-submitted ML jobs within a shared ML cluster. 
As illustrated in Figure~\ref{figure:design}, \sys's design is structured as follows:

The \i{Job Profile Layer} allows users to define job and environment requirements with a self-contained profile syntax, ensuring unambiguous execution of jobs.

The \i{Adapter Layer} is the interface between users and the cluster, facilitating job submission and monitoring.

The \i{Scheduler Layer} manages resource allocation and job scheduling, employing an optimized FCFS with a backfill mechanism to prioritize and utilize resources efficiently.

The \i{Executor Layer} is responsible for executing ML jobs on cluster nodes, featuring network file systems and customizable execution backends for scalability.

This abstraction offers a key advantage by decoupling the ML job execution workflow to independent input and output in each layer, creating a versatile design framework where different components can be combined to accomplish the cluster management objective. 

\subsection{Layer 1: Job Profile Layer} 
\label{section:design:layer1}

\begin{figure}[h!]
    \includegraphics[width=0.48\textwidth]{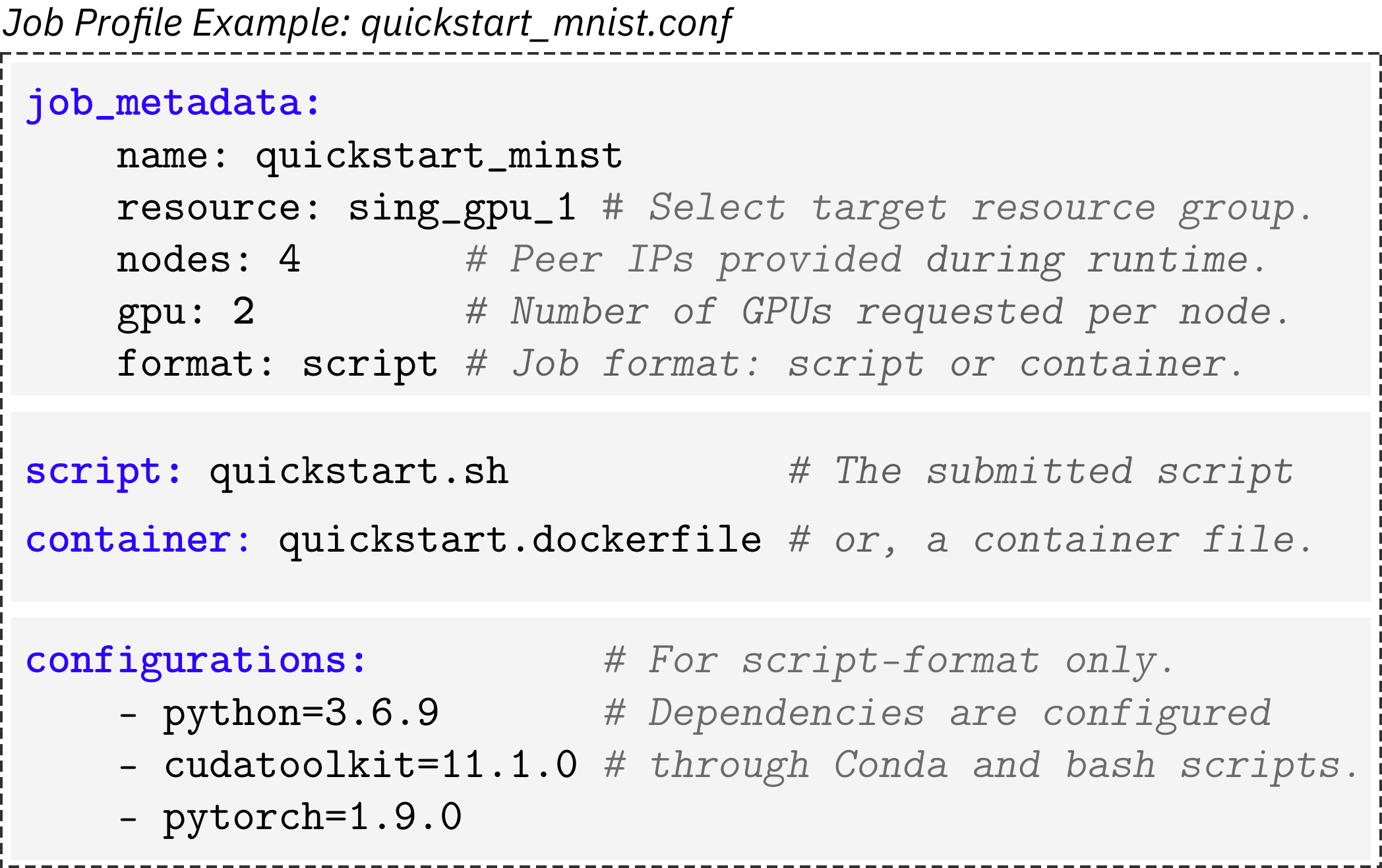}
    \caption{Example job profile where users specify job requirements, code entry point, and configurations.}
    \label{figure:layer1} 
\end{figure}

Jobs submitted to \sys are defined using job profiles, and an example showing the syntax of these profiles is in Figure~\ref{figure:layer1}. These profiles consist of the following sections:

\begin{ol}
\item Metadata and Resource Request: users specify resource requirements, including the selected resource group, number of nodes (which can be a range), the number of GPUs required on each node, and the format of the job code.
\item Job Code: this section points to the entry point of the user's job code, which can be a script or a container file.
\item Configurations: users specify the job's configurations, such as runtime libraries like CUDA version and programming dependencies like Python packages.
\end{ol}

\t{\underline{Feature \#1: Job Format Flexibility.}}
\sys supports script-format (common in HPC) and container-format job profiles simultaneously. 
Users can specify their preferred format by indicating it in the $job$ section and providing either an execution script or a container file (Dockerfile) in the $entrypoint$ section of the job profile. 
This versatility in format minimizes migration costs during job submissions across different computing environments.
Subsequently, the job profile is compiled into executable scripts by the adapter layer (\S\ref{section:design:layer2}) before being submitted to the scheduler layer (\S\ref{section:design:layer3}) for execution.

\t{Comparing to traditional HPC clusters.} 
Having a profile layer eliminates the need for setting up a server from where users submit jobs via Slurm toolchains.
In traditional HPC clusters, this server often called a \i{login node}, is a common single point of failure. Some users might unintentionally execute their job scripts directly on the login node, either for debugging purposes or due to workflow misunderstandings.
Running scripts locally on the login node can fully occupy its computing or memory resources, rendering it unresponsive and preventing other users from submitting new jobs. This can disrupt the system and potentially lead to a cluster-wide outage.
Despite warnings from HPC cluster operation teams~\cite{loginnode2, loginnode3}, such incidents still happen as they cannot be completely prevented~\cite{loginnode1}, necessitating manual recovery by the operation team which can be time-consuming. 

Another advantage of using the job profile is its self-contained and repeatable nature. 
Users can save and reuse their job profiles for recurring tasks, simplifying job submission. 
Moreover, these profiles can be shared with others to replicate experiments, fostering collaboration and reproducibility within the research community.


\subsection{Layer 2: Adapter Layer} 
\label{section:design:layer2}

\begin{figure}[t]
    \includegraphics[width=\columnwidth]{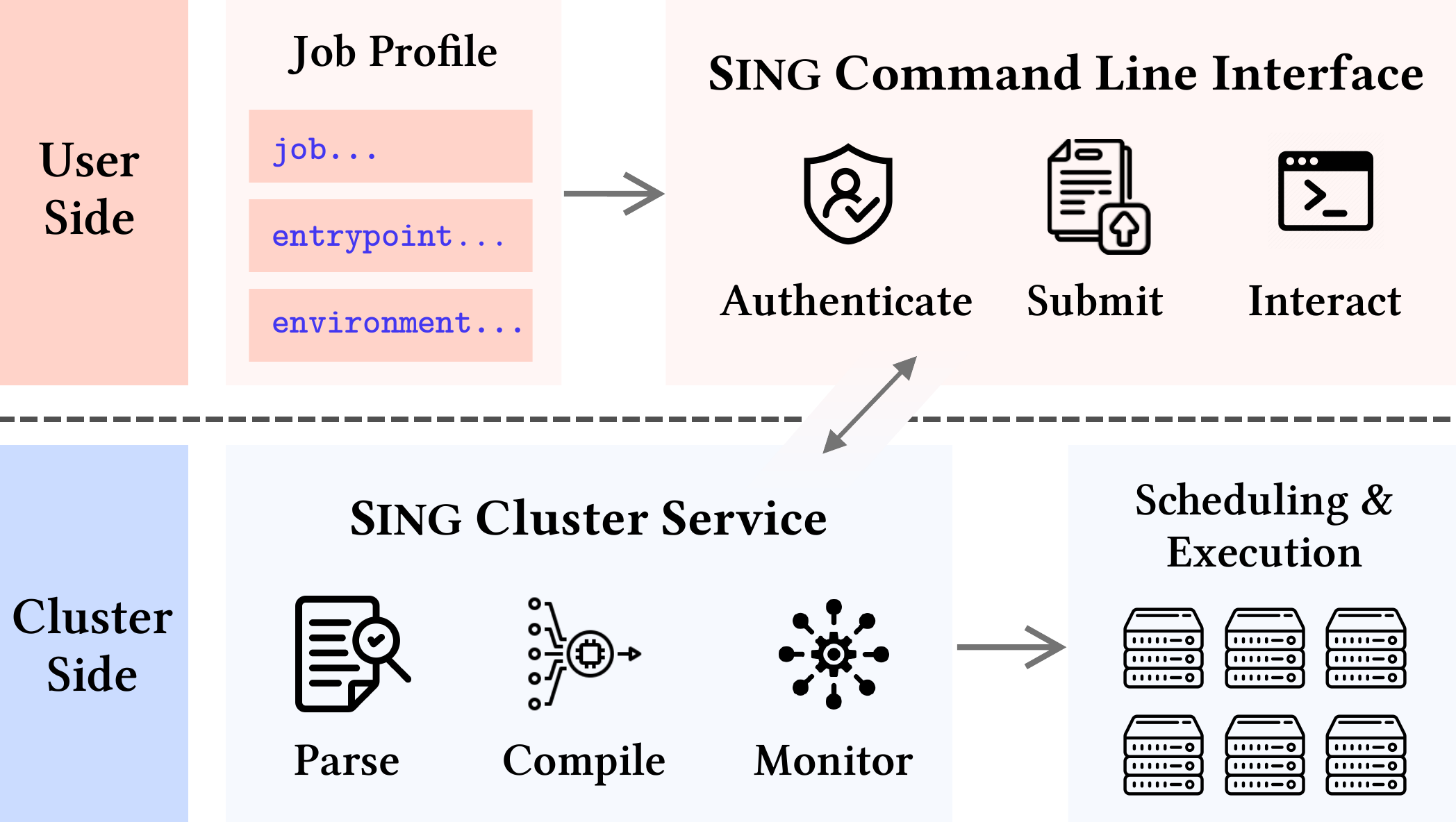}
    \caption{Workflow and functions of the adapter layer.}
    \label{figure:layer2} 
\end{figure}

The adapter layer takes the user's job profile from the previous layer and submits it to the ML cluster. It also offers user interfaces for functions like monitoring running jobs, retrieving output files, and streaming logs.

The adapter layer is referred to as such because it is programmed against the job schedulers in the layer below. It abstracts the complexity of the cluster manager's functionalities and provides a user-friendly interface for job submission and monitoring.

Our implementation, illustrated in Figure~\ref{figure:layer2}, comprises two components: \sys-cli and \sys-service, running locally on user computers and on the cluster, respectively.

\i{\sys-cli} (\sys Command Line Interface), is a local software distributed to \sys users. It stores cluster configurations, such as the cluster's IP and login credentials, and connects with the cluster. 
Beyond authentication, the \i{cli} acts as a thin layer that wraps around the services provided by \sys-service, providing commands that facilitate job submission and access to other \sys-service functionalities.

\i{\sys-service} operates on a cluster node and receives instructions from the \i{cli}. When a job is submitted, the service compiles the job files by parsing the job profile and converting it into the scheduler's recognized format. This process includes mapping configuration parameters and packaging code files before passing them to the scheduler layer.
\sys-service also provides job monitoring functions, such as fetching job status and accessing output files and logs.

\t{Connection Security.} 
\sys-cli uses remote command execution over an SSH tunnel to invoke functions in \sys-service. To authenticate users, their public key—created during registration—is added to the dedicated cluster node that runs \sys-service. This setup does not allow users to log in directly to the cluster node, thereby reducing the risk of unexpected user actions on the node that could disrupt operations~(as discussed in \S\ref{section:design:layer1}).
A more secure approach that offers better isolation between the client and cluster is to implement \sys-service functions as HTTP APIs, connected over HTTPS with password authentication. However, we opt for remote command execution as it offers more flexibility (e.g., file uploads) with less engineering effort. 
We discuss potential security concerns related to this solution in \S\ref{section:limitations}.

We discuss two key features of the adapter layer and provide details of our implementation as follows.

\f{Feature \#2: Multi-format Job Compilation.}
Job compilation in \sys enables jobs in various formats to be transformed into a standardized output script that the backend can interpret and execute. Since this compilation process is handled on the cluster side (Figure \ref{figure:layer2}), the same \sys-cli can connect to different clusters running \sys-service without requiring user code modifications. 
This design supports extensibility across various execution backends. While our \sys implementation uses Slurm~\cite{slurm} as the execution backend, community extensions (\S\ref{section:opensource}) have enabled integration with alternative backends—or even multiple backends simultaneously based on a routing strategy—without adding usability burdens on the user side.

\i{For script-format jobs}, the \sys-service identifies required dependencies and injects scripts to prepare the environment. This, for example, includes setting the \texttt{LD\_LIBRARY\_PATH} to switch between pre-installed CUDA versions~\cite{cudaversion} on cluster nodes and initializing Conda~\cite{conda} for Python dependencies. We optimize Conda initialization by caching environments using MD5 hashes of sorted dependency lists, enabling instant retrieval for recurring jobs with identical requirements.

\i{For container-format jobs}, which are self-contained packages, the job is passed to the scheduler layer for resource allocation. After that, the compute node invokes container runtimes (such as Docker or Containerd~\cite{containerd}) to execute the provided container file directly.
Optionally, script-format jobs can also be encapsulated into a container file to simplify execution, albeit with an added layer of abstraction.

After compilation, the job is submitted to the specified resource group and cluster. This layer provides extensibility by allowing the \sys-cli to connect to \sys-service on different computation clusters (varying in location, speed, or cost) \i{without} user code modification. Additionally, the \sys-service compiles the job profile into different output scripts suitable for its respective cluster.

\f{Feature \#3: Interactive Job Monitoring.}
Many research projects running on our cluster involve ongoing development and exploratory experiments. 
Users therefore need to interact with their jobs in real time, especially when debugging or monitoring the training process.
Hence, \sys provides the following interactive features.

\i{1. Remote port forwarding.} For real-time debugging tools such as Jupyter Notebook or TensorBoard, \sys-service dynamically edits port forwarding rules on the cluster gateway server, which runs HAProxy~\cite{haproxy}. This setup exposes the container's port through a randomly assigned port on the gateway. Users can then access the service using the gateway's public IP and the forwarded port. Once the job is stopped, the executor backend automatically triggers~\cite{strigger} the removal of the port forwarding rule. Note that this approach opens a public port on the gateway server, which could allow unauthorized access if the service behind the forwarded port lacks authentication. We discuss this security risk and a potential solution in~\S\ref{section:limitations}.

\i{2. Remote SSH access.} Users can gain SSH access to the remote host (either a server or a container) where their job is running to perform debugging tasks. To facilitate this connection, \sys-service generates a one-time SSH key pair and adds the public key to the authorized keys file on the remote host. The service then exposes the SSH port of the remote host on the gateway server using the previously mentioned port forwarding function. 
The private key is securely transmitted back to \sys-cli via the cli-to-service connection (\S\ref{section:design:layer2}). \sys-cli then use this private key to establish an SSH connection to the remote host through the gateway. After the job stops, both the key pair and the port forwarding rule are removed. 

\i{3. Retrieving log and output files.} \sys-cli can download or stream remote files (similar to \texttt{tail --follow})  in the job's runtime directory to access output or log.

In practice, to prevent low resource utilization resulting from remote job debugging, we restrict the usage of remote functions to jobs that require two or fewer GPUs. This encourages users to initially test their jobs with fewer resources and scale up once the testing phase is complete.


\subsection{Layer 3: Scheduler Layer} 
\label{section:design:layer3}

The scheduling layer is responsible for making decisions on the allocation of resources.
ML cluster scheduling is a popular research area~\cite{green, pollux, sia}. However, as many approaches improve cluster efficiency by evaluating jobs with performance factors such as estimated job completion time, resource utilization, and statistical efficiency, they may not be suitable for shared ML clusters in research institutions because:

\begin{ul}
\item They determine the order of job execution based on optimization algorithms targeting cluster-wide metrics. This approach may sometimes result in unfair resource allocation from the perspective of individual users.
\item They often require modifications to user code, including model hyperparameters like batch size, to exploit optimization opportunities. In the context of ML research, such modifications are less desirable, if not unacceptable, as echoed in recent literature~\cite{lucid}.
\end{ul}

\begin{figure}[t]
    \includegraphics[width=0.48\textwidth]{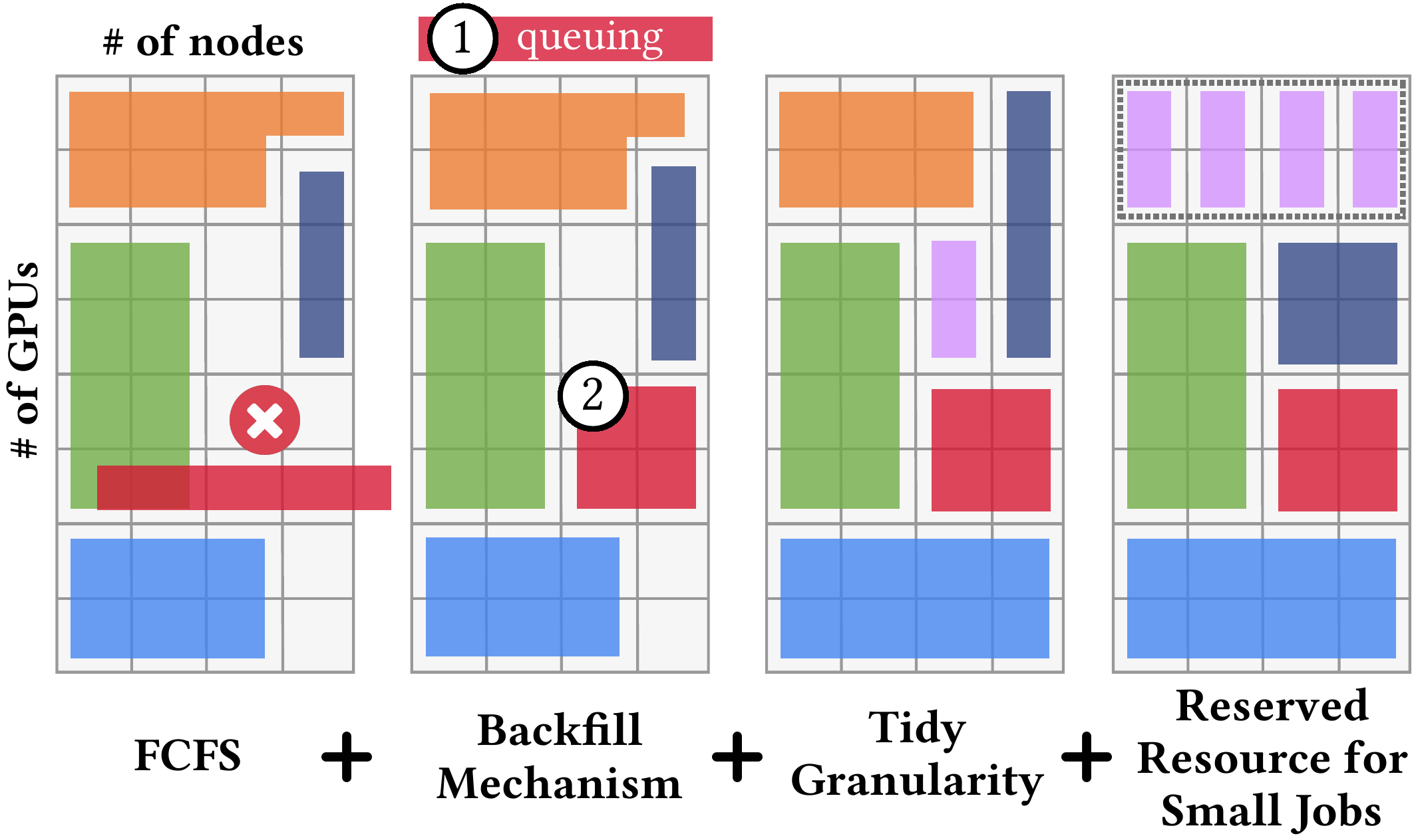}
    \caption{\sys's scheduler layer implements mechanisms to improve cluster-wide resource utilization.}
    \label{figure:layer3} 
\end{figure}

\sys incorporates the following mechanisms to improve cluster-wide resource utilization while being fair to individual users, as illustrated in Figure~\ref{figure:layer3}. Resource abuse prevention is discussed in \S\ref{section:usage:incidents}.

\f{Feature \#4: Fair FCFS with Backfill.}
\sys's scheduling approach is \i{analogous} to the way universities manage their public facilities, recognizing that ML clusters function as university-wide shared resources. Fairness in this context means that resource requests are generally processed on a first-come, first-served (FCFS) basis to ensure equal access for all users. However, to enhance overall cluster efficiency, certain requests may be given higher priority than earlier ones, as detailed below, in order to improve cluster-wide resource utilization. 

\sys uses a scheduling strategy that combines optimized FCFS with a backfill mechanism, as depicted in Figure~\ref{figure:layer3}. When a job is submitted, it is placed in a queue, awaiting execution. 
The scheduler engages in a bin-packing process to determine if there are sufficient available resources to start the next job.
In cases where the available resources cannot satisfy the first job in the queue, the backfill mechanism will help avoid head-of-line blocking due to large jobs in the front of the queue: it temporally skips the first job and evaluates the subsequent jobs to identify the ones that can be executed with available resources.

Next, we introduce strategies that further enhance cluster utilization by minimizing fragmentation and prioritizing smaller jobs. The usage statistics in \S\ref{section:usage} validate the effectiveness of these strategies.

\t{Threshold for Backfilled Requests.}
In some cases, the backfill mechanism may lead to the starvation of the first job, especially if the first job request is large while the subsequent request is small. To address this, we restrict the total number of allocated GPUs for backfilled requests with a threshold (in \sys, we set the threshold to the size of the first job request).
Once this limit is reached, priority is given to the first job to make sure it will be scheduled next.
Algorithm~\ref{algorithm} illustrates the pseudocode of this scheduling strategy.

\t{Tidy Granularity.} 
\sys accepts two-dimensional resource requests: the number of nodes and the number of GPUs per node. To prevent fragmentation in the bin-packing process, we have established a limited granularity for the quantity of GPUs in each node type, offering preset choices of 1/2/4/8-GPU nodes. Additionally, when users request multiple nodes, they must select nodes of the same type.
These predefined granularities, when carefully scheduled with bin-packing algorithms~\cite{binpacking}, can enhance resource utilization as they minimize small fragments during scheduling.
From a user perspective, these choices of granularities align well with typical ML experiment settings (i.e., powers of two).
The allocation of CPU and memory is proportional to the number of GPUs in each node type, which facilitates data processing during GPU training. This approach is generally sufficient based on our operational experience, as our GPU nodes are equipped with 40-core CPUs and 256GB of memory (which exceeds the total GPU memory on a node). For tasks that require only CPU or memory resources, we recommend using CPU clusters.

\begin{algorithm}[t]
    \DontPrintSemicolon
    \SetKwInput{KwData}{Input}
    \KwData{
    \begin{ul}
    \item Job Queue $\mathcal{Q}$, Backfill Threshold $\mathcal{T}$
    \end{ul}
 }
    \Begin{
        \tcc{Decide the next job to run}

        $job$ = $\mathcal{Q}$.peek() \\

        \tcc{Check the first job in queue }
        \If{available(job)}{ 
            $\mathcal{Q}$.pop() \\
            $\mathcal{T}$ = $\mathcal{Q}$.peek().GPUs \\
            \b{return} job \\
 }

        \tcc{Try backfilling jobs}
        \For{backfill\_job \b{in} $\mathcal{Q}$}{
               \If{available(backfill\_job) \b{and} $\mathcal{T} \ge \textit{backfill\_job}\mathrm{.GPUs}$ }{
                    \tcc{Reduce backfill threshold}
                    $\mathcal{T}$ = $\mathcal{T}$ - \textit{backfill\_job}.GPUs \\
                    \b{return} \textit{backfill\_job}
 }
 }
        \b{end}\\
 }{\b{end}}
    \caption{Optimized FCFS with Backfill}
    \label{algorithm}
\end{algorithm}

\t{Reserved Resource for Small Jobs.}
Another effective strategy we implemented is the prioritization for smaller jobs~(illustrated in Figure~\ref{figure:layer3}). We partition some 8-GPU nodes into 1/2/4-GPU nodes with isolated computation and network environments with SR-IOV~\cite{sriov}, which allows a single physical NIC to appear as multiple virtual NICs with minimal performance loss~\cite{sriov-ms}. These nodes are exclusively allocated to serve small requests. 
This strategy offers priority incentives that encourage users to request fewer resources when they only need to debug or quickly test for a few epochs, reducing the waiting time for other users.
This optimization also implicitly resembles the effects of a short-job-first strategy, which reduces the waiting time for queueing jobs. 

\f{Feature \#5: Early-Initialization and Caching}.
When jobs are queued and awaiting execution, \sys accelerates the bootstrap stage by \i{pre-initializing} Conda environments (for script-format jobs) or building container images (for container-format jobs) as immutable directories stored in distributed network storage. This network storage is accessible to all nodes (as described in Feature \#6 in \S\ref{section:design:layer4}).

To further reduce job startup time, \sys caches the dependency environment and container images locally on each node. When a job is started on compute nodes, \sys first checks if the resource is available without accessing the nodes' network storage. If not, the environment is fetched to the local node for job execution and cached there for subsequent executions, unless purged due to local storage space constraints under the LRU cache replacement policy.

\begin{figure}[b]
    \includegraphics[width=\columnwidth]{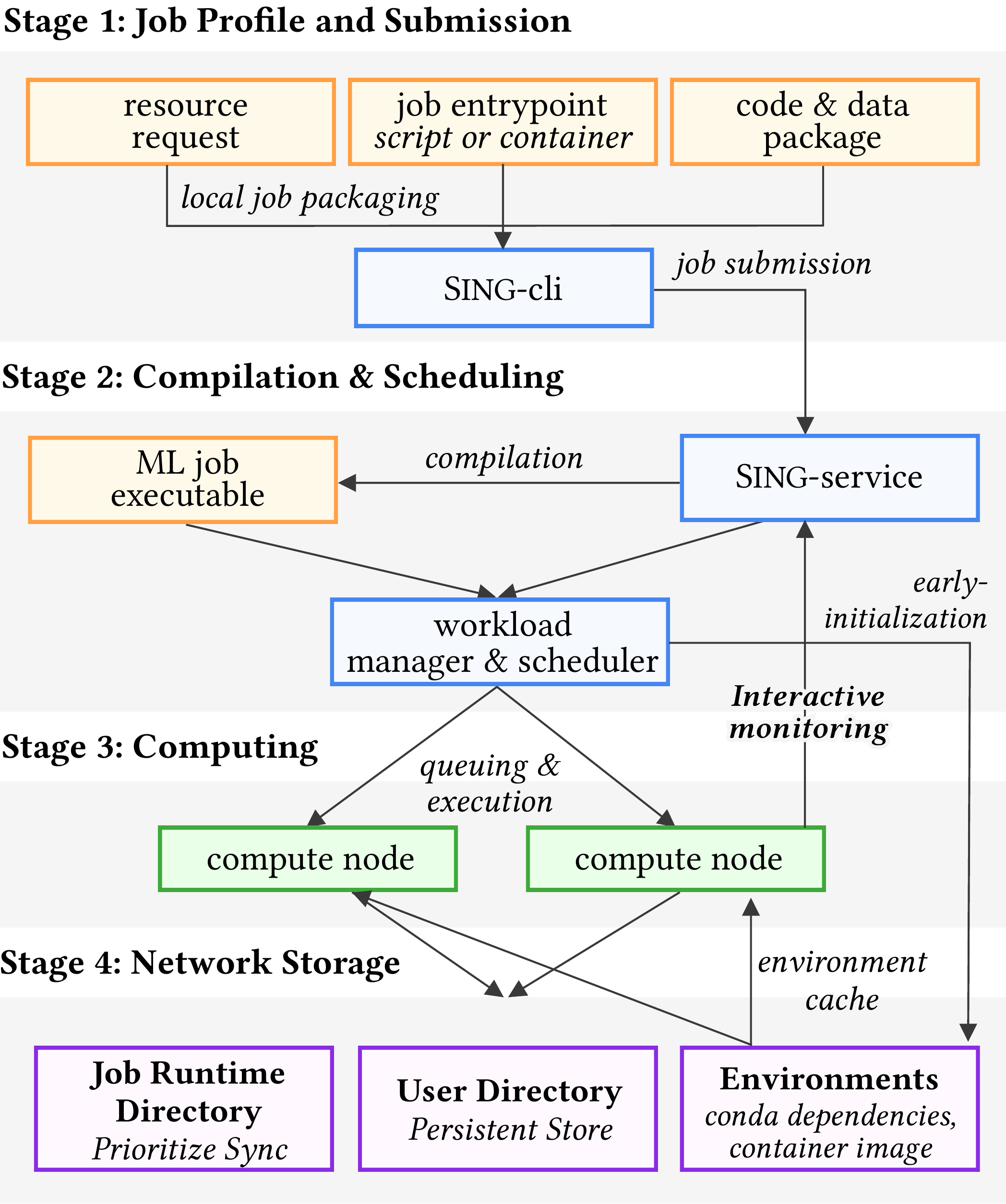}
    \caption{An end-to-end workflow of job processing in \sys, from job submission to execution.}
    \label{figure:layer4} 
\end{figure}

\subsection{Layer 4: Execution Layer}
\label{section:design:layer4}

The executor layer is where job profiles become instances and bootstrap for execution, and it provides necessary architecture support for parallel computation and resource isolations for multi-job processing.
Figure~\ref{figure:layer4} illustrates the topology and architecture of  \sys that supports distributed job execution. 
\sys utilizes Slurm~\cite{slurm} as the default backend and this section discuss its capability and our enhancements to Slurm to make it a better fit for ML cluster management. 
\S\ref{section:opensource} discusses a community work that utilizes Kubernetes for executor backend in \sys's architecture.
\S\ref{section:discussion} discusses the reason why we choose Slurm as default backend.

\f{Feature \#6: Distributed Job Bootstrapping.}
Each cluster node runs a Slurm agent that manages job processes. For parallel jobs spanning multiple nodes, an environment variable with assigned IPs facilitates peer connections. The agent executes the compiled job script from the job profile in the adapter layer (in \S\ref{section:design:layer2}), starting the initialized environment~(Conda or container) and calling the job script within it. This environment is pre-initialized and cached in the network file system, as described in Feature \#5 (\S\ref{section:design:layer3}).

We deploy a network file system with RDMA support using GlusterFS~\cite{glusterfs} for file synchronization across nodes and to prevent file duplication. 
There are two user-level file spaces: the job runtime directory and the user directory. Both directories are accessible during job runtime or using \sys-cli but offer different Quality of Service (QoS) guarantees. 
The user directory provides persistent storage shared across multiple job submissions, prioritizing stable storage over synchronization speed. In contrast, the job runtime directory offers faster storage that only persists for a short period after job completion unless the user submits a new job. Otherwise, it is automatically purged.

\f{Feature \#7: Heterogeneous Executor Backends.} 
\sys's executor backend, Slurm, operates under a "controller-agent" model. In this setup, the controller acts as a puppet master, managing resource groups, while agents run on compute nodes to execute jobs. The controller can partition resources into different groups (e.g., CPU-only, consumer GPU, datacenter GPU), allowing users to specify the cluster and optionally the resource group to meet heterogeneous hardware requirements, while \sys enforces user resource quotas.

Additionally, resources can be added or removed from the cluster simply by starting or stopping the agent. This model enables dynamic scaling of resources across the cluster, allowing it to adapt to varying workloads and user demands. Additionally, cloud-based compute nodes can be scaled through an agent running in the cloud, which connects back to the controller. 
We utilize a self-built cloud instance image (e.g., Amazon Machine Images~\cite{ami}) that embeds the agent and starts it automatically at instance boot, to simplify the process. This approach ensures scalability and flexibility, adding additional resources to the cluster while remaining transparent to users. 
Future implementations may fully automate scaling operations based on cluster occupancy and cloud resource budgets.

\vspace{-10pt}
\subsection{\sys Implementation} 

We implement the scheduling and execution layer of \sys based on Slurm~\cite{slurm}, which is extended to support the scheduling policies in \S\ref{section:design:layer3}. 
We selected Slurm for its popularity and simplicity in academic cluster management (further discussion in \S\ref{section:discussion}). We note that Kubernetes~\cite{k8s} can provide similar functionality and can also serve as the executor backend for \sys, which has been developed in the Kubernetes community and is discussed in \S\ref{section:opensource}.

\section{Usability and Accessibility}
\label{section:design:analysis}

This section discusses the usability of \sys's design from the perspective of cluster operators and users, reflected by its operational workload and user experience.

\subsection{Operation and Maintenance Cost}

Table~\ref{table:cost} compares the estimated human effort needed to build \sys from scratch (17 person-days), or adopt the open-source \sys configuration and code (8 person-days). 
The difference is mainly due to \sys-cli and \sys-service development and figuring out software configurations. See our open-source plan in \S\ref{section:usage:trace}.

\t{Installing System Software.} 
The effort required for the initial deployment varies depending on the operation team's familiarity with Linux software configurations. For instance, setting up GlusterFS for the networked file system and Slurm for the scheduler is relatively straightforward if the team efficiently understands and follows their well-documented setup procedures. Meanwhile, they also have a large community where solutions to most deployment issues can be readily found.


\t{Service Requests and Maintenance.}
The operation team's day-to-day workload is primarily on responding to user requests and maintaining the cluster's stability: 
\begin{ul}
\item To reduce basic service requests, our team provide clear job examples under various distribution settings and programming frameworks, helping users quickly get started with these templates. We also outlines detailed steps and best practices for job debugging. As a result, over the past 12 months, our cluster has seen an average of one service request for every 940 jobs submitted.
\item System failures in \sys are rare due to the minimalism in software architecture, which minimizes the risks of software failure. Hardware and networking issues, such as network file system delays, are discussed in~\S\ref{section:usage}.

\end{ul}

\begin{table}[t]
    \caption{Breakdown of estimated human efforts for building the \sys solution in a new cluster.}
    \centering
    \small
    \begin{tabular}{l c c}
    \toprule
    \multirow{2}{*}{\b{Tasks / Steps}} & \multicolumn{2}{c}{\makecell{\b{Est. Person-Days}}}  \\ \cline{2-3} \\[-8pt]
    & Ground-up & Adopt 
    \\[2pt] \hline \\[-8pt]
 1. Basic OS and GPU Driver Configs & 2  & 2 \\[2pt] \hline \\[-8pt]
 2. Cluster Softwares: Slurm and NFS & 4 & 3 \\[2pt] \hline \\[-8pt] 
 3. Integrate \sys-cli and -service & 8 & 2 \\[2pt] \hline \\[-8pt]
 4. Testing \& Documentation & 3  & 1 \\
    \bottomrule
    \end{tabular}
    \label{table:cost}
\end{table}

\subsection{User Experience}

Our design strives to improve users experience for ML clusters, yielding several key advantages, as discussed below.

\t{User-Friendly Features.} \sys-cli offers a suite of functions tailored to typical ML research workflows, allowing users to easily access remote resources. These functions include remote file uploads/downloads, attaching bash environment, and remote port forwarding.

\t{Simplified Environment Management.} Users no longer need to maintain two sets of scripts for local and remote executions. The job profile encapsulates all environment dependencies as well as scripts to start the ML process, and the steps to execute jobs locally and remotely are identical. 

\t{Documentation.} 
The user documentation for \sys offers a detailed collection of clear and intuitive examples for job submissions, ranging from basic workflow testing to hybrid parallel training. These examples help users to quickly understand the interface and functionality of \sys, providing easy-to-follow solutions while  reducing the operational team's workload in addressing user inquiries.


\vspace{3pt}
\t{Summary.} Cluster operation includes both problem \i{prevention} and \i{resolution}. We prioritize \i{prevention} by adhering to design principles that reduce architectural complexity and use foolproof interfaces. For \i{resolution}, please see \S\ref{section:usage:incidents}.

\section{Operation and Analysis} 
\label{section:usage}

We analyze \sys's performance and usage over~2.5 years to offer insights into academic ML cluster operation. The cluster consists of 128 Nvidia GTX 3090 GPUs on 16 nodes.

We first analyze cluster-wide trends including temporal patterns~(\S\ref{section:usage:pattern}) and job-specific characteristics~(\S\ref{section:usage:job}). 
Then, we discuss the experiences and actions taken to handle resource bottlenecks and hardware failures (\S\ref{section:usage:incidents}). 
Finally, we present the schema of the \sys cluster trace, a more comprehensive trace compared to prior work in \S\ref{section:usage:trace}.

\subsection{Usage Statistics: Temporal Patterns}
\label{section:usage:pattern}



\sys exhibited various temporal patterns during operation. These patterns are instrumental in planning for capacity and system maintenance schedules. 
Unless otherwise specified, the period under review in this section is for the 30 months from July 2021 to December 2023.

\begin{figure}[t]
    \centering
         \includegraphics[width=\linewidth]{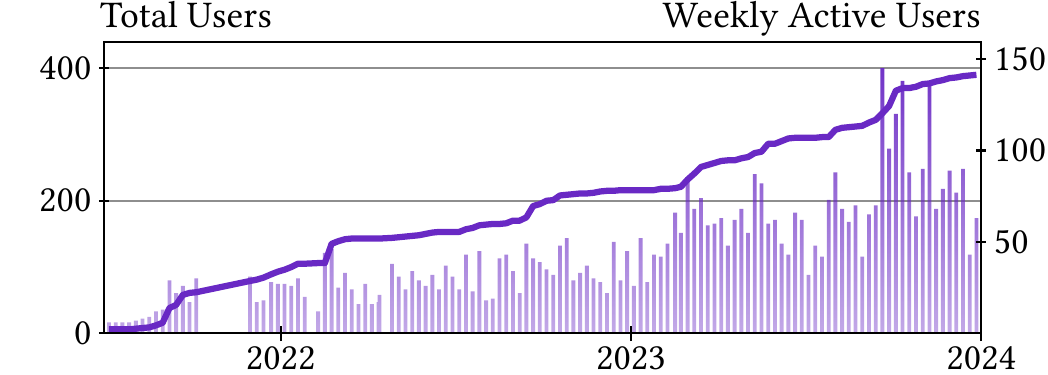}
         \vspace{-20pt}
         \caption{User growth and weekly active users.}
         \label{figure:usage:user}
 \end{figure}

\t{Number of Users.} 
Figure~\ref{figure:usage:user} shows periodic surges in new user registrations for \sys, where spikes are visible around September and February, aligning with the distribution of email newsletters at the beginning of each semester to introduce new researchers to the platforms. 
Outside these periods, user growth is steady and gradual, attributed to word-of-mouth referrals. Additionally, the data reflects an increasing trend in new user registrations over time, showing the growing popularity of the \sys within our institution.

\begin{figure}[t]
     \centering
      \begin{subfigure}[b]{\linewidth}
         \includegraphics[width=\linewidth]{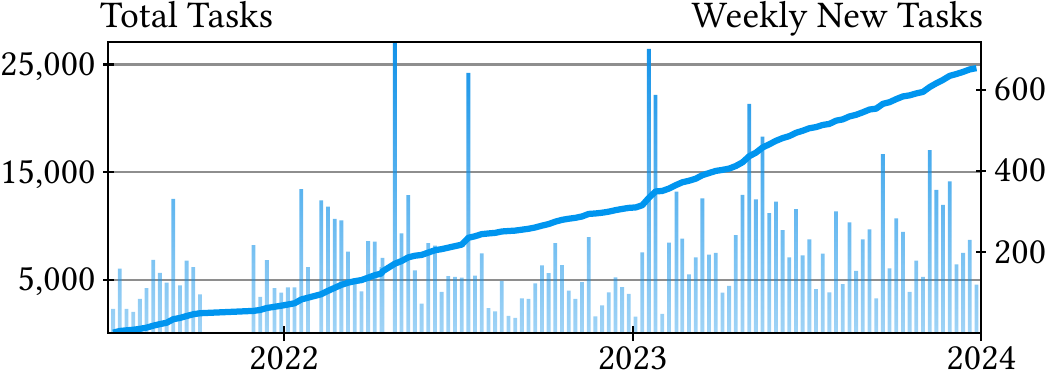}
         \vspace{-10pt}
         \caption{Number of total tasks and weekly new tasks from 2021 to 2023.}
         \label{figure:usage:task}
    \end{subfigure}

    \begin{subfigure}[b]{0.59\linewidth}
         \includegraphics[width=\textwidth]{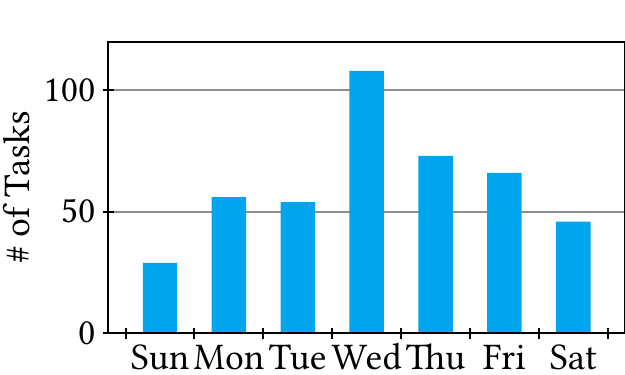}
         \captionsetup{width=0.95\linewidth}
         \caption{Tasks over a week}
         \label{figure:usage:taskdow}
     \end{subfigure}
     \hfill
     \begin{subfigure}[b]{0.40\linewidth}
         \includegraphics[width=\textwidth]{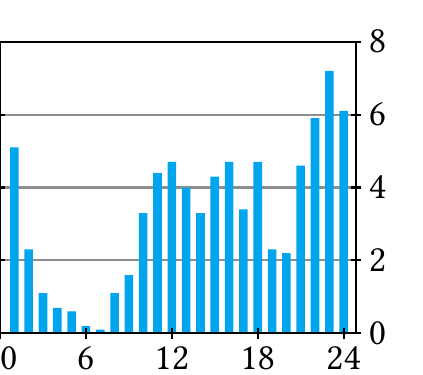}
         \captionsetup{width=\linewidth}
         \caption{Tasks over a day}
         \label{figure:usage:taskhod}
     \end{subfigure}
     \vspace{-20pt}
     \caption{Task submissions over a year in monthly, daily, and hourly submission patterns.}
     \label{figure:usage:taskall}
 \end{figure}

\t{Task Submissions.}
Figure~\ref{figure:usage:task} shows the task submission statistics of \sys, demonstrating a consistent growth in task submissions, with over 200 weekly average submissions. Notable submission spikes are observed around January 2022, late April to early May 2022, and late January 2023, possibly corresponding to periods with paper submission deadlines.

Figure~\ref{figure:usage:taskdow} shows a mid-week peak in task submissions, particularly on Wednesdays, and gradually decreases towards the weekends, with the lowest on Sundays.
Figure~\ref{figure:usage:taskhod} illustrates a daily trend where submissions are minimal in the early morning, gradually increasing and peaking in late evenings, especially between 10pm through 1am the next day. 
This trend may reflect researchers' late-night work habits and highlights the importance of a low-maintenance cluster, given the limited availability of operators during late hours.

\begin{figure}[t]
    \includegraphics[width=\linewidth]{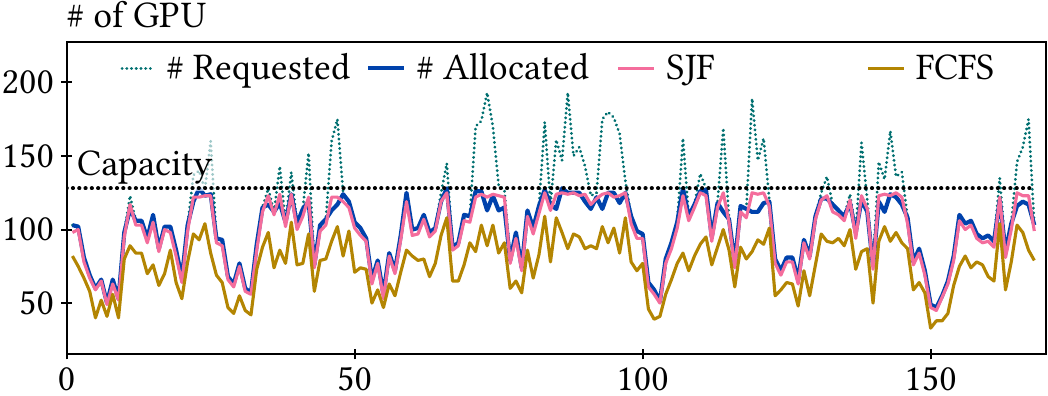}
    \vspace{-16pt}
    \caption{GPU occupancy over a week, from Sunday 0am to Saturday 11pm (168 hours).}
    \label{figure:usage:gpu}
    \vspace{10pt}
    \includegraphics[width=\linewidth]{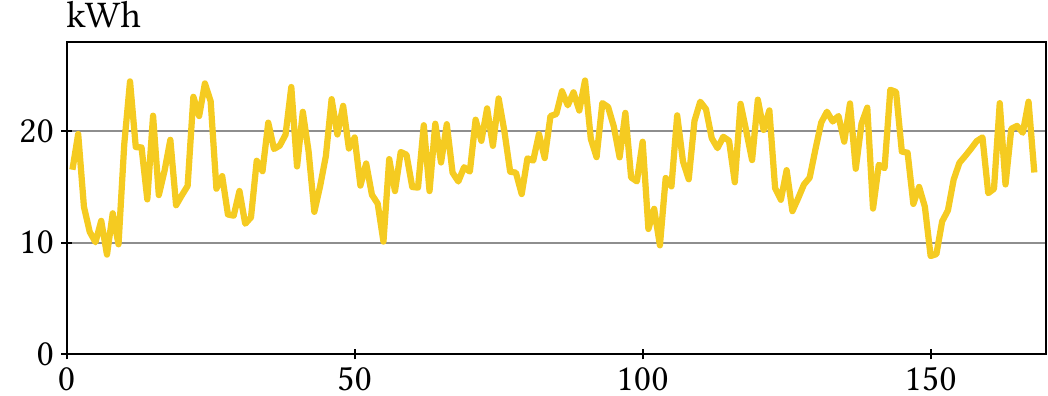}
    \vspace{-16pt}
    \caption{Cluster-wide power draw over the same time period as in Figure~\ref{figure:usage:gpu}, showing the GPU power is dominant.}
    \label{figure:usage:power}
\end{figure}

\t{GPU Occupancy Patterns.} 
Figure~\ref{figure:usage:gpu} illustrates the GPU occupancy rate over a busy week, with the x-axis representing hours from 0 to 167, corresponding to Sunday 0am to Saturday 11pm.
During the sampled period, \sys experienced job oversubscription for about 30\% of the time, with peaks lasting around two hours. 
We calculate the GPU assignment rate as the sum of assigned GPUs at each hour divided by the sum of the minimum value between the number of requested GPUs and the cluster capacity at each hour. This approach reflects utilization adjusted for demand.
\sys maintained a GPU assignment rate of over 96\%, reflecting the effectiveness of our backfilling scheduling mechanism (\S\ref{section:design:layer3}). 
Less than 4\% of GPUs are idle during the oversubscription period due to resource fragmentation, optimized by pre-defined granularity and the bin-packing algorithm (\S\ref{section:design:layer3}).

\t{Comparing with Alternative Scheduling Policies.} 
Using the job trace shown in Figure \ref{figure:usage:gpu}, we conducted a mini-evaluation of alternative scheduling policies to determine job priorities and compare their performance with \sys. 
We replay the job trace in a simulator. The simulator functions as a step-based discrete-time system with a 1-minute granularity, resulting in 10,080 steps for this evaluation. Each step performs three tasks: 
\begin{ol} 
    \item Calculate job progress to identify completions. 
    \item Check the job trace and add new arrivals to the queue. 
    \item Apply scheduling policy to start new jobs. 
\end{ol}

We compared two baseline policies: (1) the First-Come, First-Served (FCFS) policy, which is the default in Slurm, and (2) the Short-Job-First (SJF) policy used in Lucid~\cite{lucid}, where we provided the exact running time to simulate an ideal scenario. 
Like in \sys, this evaluation did not include job resource scaling or preemption. Additionally, all jobs were subject to a 24-hour rule (\S\ref{section:usage:incidents}), consistent with the policy applied in \sys from which the job trace was collected.

For the evaluation metrics, we examined job occupancy rates and queueing times. The GPU assignment times for \sys, FCFS, and SJF were 96\%, 75.1\%, and 96.5\%, respectively, as illustrated in Figure \ref{figure:usage:gpu}. The average queueing times for \sys, FCFS, and SJF were 0.81, 1.32, and 0.64 hours, respectively. \sys outperformed FCFS due to its backfill mechanism, while SJF performed better than \sys in reducing waiting times (note that \sys's dedicated resource pool for small jobs contributes to the low average queueing time). However, in terms of tail (maximum) queueing times, \sys, FCFS, and SJF recorded 7.2, 7.4, and 10.2 hours, respectively, indicating that SJF may lead to starvation for longer jobs.

\t{Power Consumption Metrics. }
Figure~\ref{figure:usage:power} illustrates the power consumption over the same time frame as the GPU occupancy graph. Max power draw is crucial for cluster planning. 
The primary reasons for the remaining differences are as follows: (1) the base power draw, which includes the idle power of the CPU, GPU, and other components, is relatively constant and independent of the GPU occupancy rate; and (2) the GPU utilization rate also influences power draw~\cite{zeus}, meaning that two jobs occupying the same number of GPUs may exhibit different power consumption if one job is more computationally intensive than the other.

\subsection{Usage Statistics: Job-level Characteristics}
\label{section:usage:job}

We present job-level characteristics of how long jobs run, how long they wait in the queue, and how efficiently they use resources, collected during the operation of \sys.

\begin{figure}[b]
    \centering
     \begin{subfigure}[b]{\linewidth}
        \includegraphics[width=\linewidth]{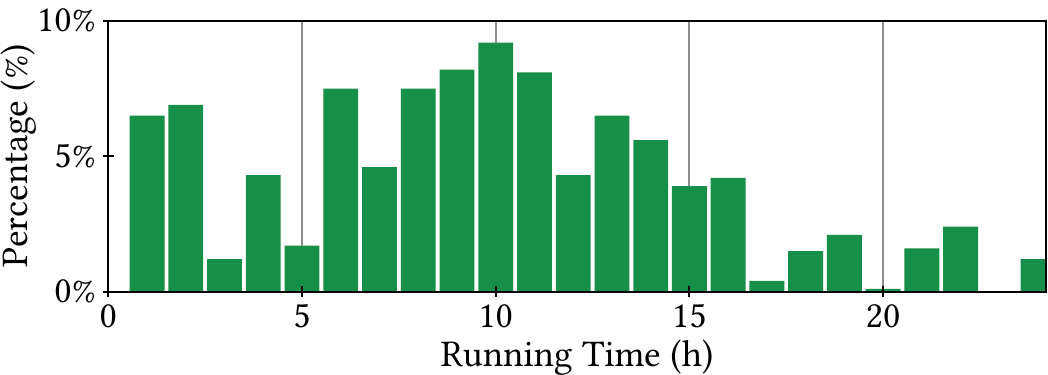}
        \caption{Distribution of job run-times in the GPU cluster, with significant clusters observed at 1-2 hours and 8-11 hours.}
        \label{figure:usage:run}
    \end{subfigure}
    
     \begin{subfigure}[b]{0.513\linewidth}
        \includegraphics[width=\textwidth]{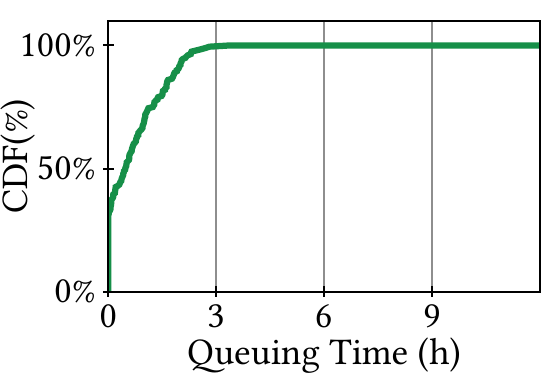}
        \captionsetup{width=0.95\linewidth}
        \caption{Queuing-Time Distribution}
        \label{figure:job:queue1}
    \end{subfigure}
    \hfill
    \begin{subfigure}[b]{0.477\linewidth}
        \includegraphics[width=\textwidth]{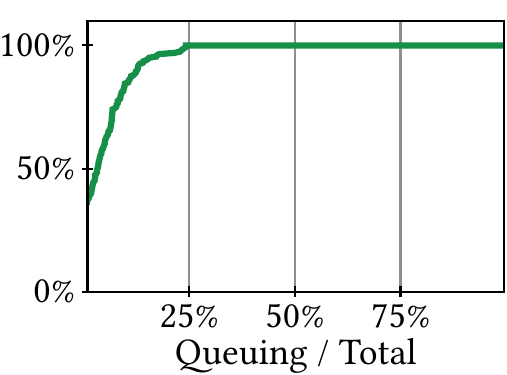}
        \captionsetup{width=\linewidth}
        \caption{Queuing / Run-Time \%}
        \label{figure:job:queue2}
    \end{subfigure}
    \vspace{-20pt}
    \caption{Analysis of job run-times (a) and queuing times (b and c) in the GPU cluster, illustrating the distribution of job durations and queuing times.}
\end{figure}

\t{Job Running-Time.} 
Figure~\ref{figure:usage:run} illustrates the job durations within the GPU cluster. The majority of job lengths fall below 2 hours or between 6 to 11 hours. This pattern is a result of both user preferences and our scheduling approach.
As discussed in \S\ref{section:design:layer2}, \sys supports interactive usage of cluster resources, such as port forwarding and remote bash access, but restricts these features to smaller job requests. This encourages users to initially debug their jobs with smaller resource requests. 
During peak hours, \sys automatically imposes a maximum runtime of 24 hours \i{per submission} to ensure fair usage (\S\ref{section:usage:incidents}). Users with longer-duration needs are advised to checkpoint their jobs around the 12-hour mark and submit a new job to continue after checkpointing. 
These mechanisms explain the prevalence of jobs running for less than 2 hours and between 6 to 11 hours.

\t{Job Queuing-Time.} 
Figures~\ref{figure:job:queue1} and \ref{figure:job:queue2} displays the cumulative distribution function (CDF) of job queuing times in hours and the queuing time as a percentage of the total job time. These figures highlight that over 40\% of jobs have no waiting time, and more than 90\% of jobs wait less than 1 hour or 10\% of their total runtime.
These statistics are influenced by \sys's design. Short jobs receive prioritized allocation with dedicated resource and backfilling mechanism~\cite{backfilling} to improve cluster-wide efficiency. Longer jobs are required to be checkpointed within 24 hours and resubmitted to prevent excessive occupation during peak hours.

\begin{figure}[b]
        \includegraphics[width=\linewidth]{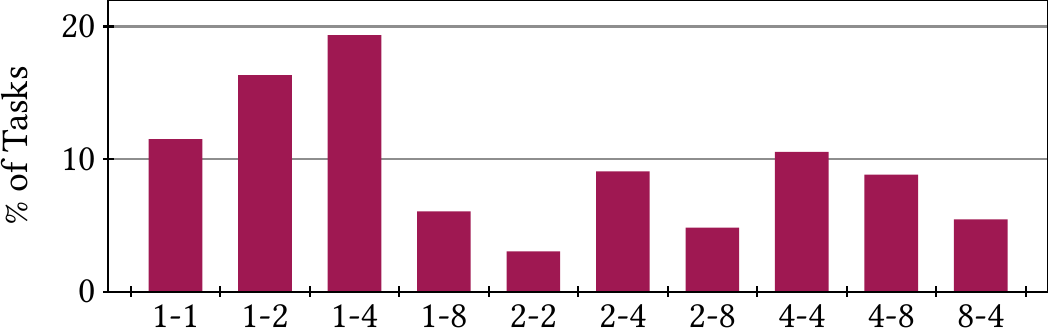}
        \caption{Distribution of job sizes, measured by the number of GPUs requested (e.g., 2-4 is two nodes with 4 GPUs each).}
        \label{figure:job:size}
\end{figure}

\begin{figure}[b]
     \begin{subfigure}[b]{0.511\linewidth}
        \includegraphics[width=\textwidth]{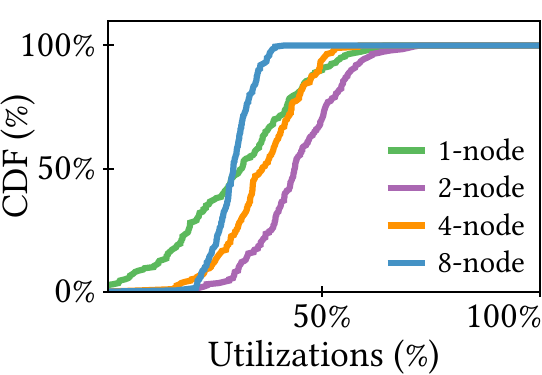}
        \captionsetup{width=0.95\linewidth}
        \caption{GPU Utilization}
    \end{subfigure}
    \hfill
    \begin{subfigure}[b]{0.478\linewidth}
        \includegraphics[width=\textwidth]{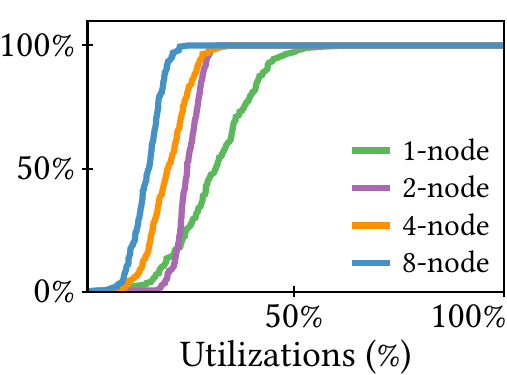}
        \captionsetup{width=\linewidth}
        \caption{GPU Memory Util.}
    \end{subfigure}
    \caption{CDFs of job-level GPU and memory utilization, illustrating resource consumption across various job sizes (i.e., number of nodes requested).}
    \label{figure:job:util}
\end{figure}
 
\t{Job Sizes Distribution.}
Figure~\ref{figure:job:size} shows the top 10 job sizes by resource requests, with about 40\% being small jobs requesting fewer than four GPUs. This is influenced by \sys's design, which favors smaller jobs by prioritizing their execution (\S\ref{section:design:layer3}) and limits interactive debugging tools (\S\ref{section:design:layer2}) to only small jobs. 
This approach motivates users to start with smaller resource requests for initial testing and debugging, enhancing overall resource utilization. 
Additionally, we make note that job sizes are restricted to powers of two~(\S\ref{section:design:layer3}). This restriction aids in reducing resource fragmentation in the scheduling's bin-packing process and does not conflict with typical ML research experiment settings.

\t{GPU Utilization and Memory Utilization.}
Figure~\ref{figure:job:util} illustrates GPU and GPU memory utilization based on the number of nodes requested by jobs. The utilization statistics only account for periods when a GPU is occupied.
Generally, as the number of nodes in a job increases, its utilization decreases, suggesting possible synchronization overheads. 
For 8-node jobs, the average GPU utilization is 25\% and the average GPU memory utilization is 15\%. This may indicate that communication-efficient training methods may not be widely adopted in current ML experiments.

\vspace{-8pt}
\begin{table}[h!]
    \small
    \centering
    \caption{Summary of issues and types of solutions.}
    \vspace{-8pt}
    \begin{tabular}{@{\extracolsep{0pt}} r c c}
    \toprule[0.8pt]
    \b{Type} & \b{System Health Issues} & \b{Type of Solutions} \\ 
    \\[-10pt]
    \hline \\[-8pt]
 Resource & Excessive Storage Usage  & Policy Enforcement  \\[2pt] \hline \\[-8pt]
    
 Resource & Low GPU Utilization   & User Guidance  \\[2pt] \hline \\[-8pt]
    
 Resource & Long Job Running Time  & Policy Enforcement  \\[2pt] \hline \\[-8pt]
    
 Failure & Network File System Delay & Incidents Handling  \\[2pt] \hline \\[-8pt]
    
 Failure & System Configuration Lost  & Incidents Handling \\
    \bottomrule
    \end{tabular}
    \label{table:incidents}
\end{table}

\vspace{-20pt}
\subsection{System Availability \& Maintenance}
\label{section:usage:incidents}

This section shares our operational experiences with \sys.
In Table~\ref{table:incidents}, we summarize the actions that address availability challenges into three types: user guidance, policy enforcement, and specific incident handling.

\subsubsection{Resource Bottlenecks}

Running into resource bottlenecks and limitations is the top feedback we hear from cluster users. We employ the following mechanism to promote fair usage of resources.

\t{Efficient Storage Usage.}
 Training ML models often involves accessing large model parameters and datasets. While we have over 1PB storage space available, the rapid accumulation of files from a large user base can exhaust our disk space. To ensure efficient storage utilization, we have implemented the following guidance and policies:
\begin{ul}
    \item \i{Public Dataset Directory}. In addition to individual user directories, we create a public directory where commonly used large datasets~\cite{huggingface_dataset} are stored. Users can soft link these datasets directly to their directories, removing the need for duplication in downloads or copies. This approach not only prevents duplicate files but also saves download time and bandwidth.
    \item \i{File Expiration Policy}. Some users store large files in their directories that are not immediately necessary for their job. To free up storage space and promote responsible resource usage, we implement a file expiration policy: files that have not been accessed for more than 60 days are subject to deletion. To prevent accidental data loss, this expiration process is not real-time: We initiate the removal process at the beginning of each month and send email notifications seven days prior to deletion. 
\end{ul}

\t{GPU Utilization Monitoring.}
We closely monitor GPU utilization in \sys to identify cases of underutilized resources, which are often caused by non-computation-intensive tasks, inefficient job implementations, or interactive usage (only allowed on small nodes). 
Although we do not immediately intervene, we collect data and send email notifications to users with significantly low GPU utilization, and also offer help for job optimization when needed.

\t{Job Running Time Limit.}
In \sys, we limit job running time to 24 hours during peak times, defined as when there is a pending job queued for over an hour (other definitions may apply here). Users requiring longer runtimes should use checkpoints and resubmit their jobs. Experienced users can automate this process for continuous execution. 

This policy is also \i{analogous} to the way universities manage their public resources, similar to borrowing a library book: you need to return it after a certain period, but you can re-borrow if it is not reserved by someone else.
The policy is in place to prevent excessively long or unattended jobs and ensure fair allocation, particularly during peak periods.

\subsubsection{Incidents Handling}
We share lessons and recommendations from unexpected incidents.

\t{Network File System Delay.}
We observed substantial delays and inconsistencies in the network file system during peak periods of concurrent remote I/O, impacting job initialization and causing runtime errors. 
We addressed this issue by switching to RDMA connections for the network file system. Users are also advised to utilize collective communications directly over the network and avoid file synchronization during job execution.

\t{System Configuration Lost Due to Hardware Upgrade.}
While upgrading the physical network interface cards (NICs) in \sys, we encountered an issue due to the tight coupling of the container's PCI-E setting and specific NICs. 
When we switched NICs, it triggered a failure that rendered the container inoperable. 
We could not directly modify the PCI-E setting inside of the container to use the new NIC because the container was not able to start at all. 
In the end, we spent about 3 days (as per Table~\ref{table:cost}) recreating the environment. 


\section{\sys Open Source}
\label{section:usage:trace}
\label{section:opensource}

This paper represents our ongoing effort aimed at improving the management of shared ML clusters in academic institutions. 
Since \sys is currently deployed in a production cluster, we are unable to fully explore all potential design choices and optimization strategies.

To facilitate the research and operations of similar infrastructures, we provide \sys's source code, configurations, and our cluster's job traces as open-source resources~\cite{opensource}.
Researchers interested in hands-on exploration of \sys are welcome to access our on-premise deployment hosted at~\cite{tacc}.

\t{Job Trace.} The \sys trace offers more detailed job characteristics and runtime measurements compared to previous traces like Microsoft Philly~\cite{philly} and Alibaba PAI~\cite{mlaas}.

\begin{ul}
    \item \b{Job Submission History (36 months)}: 
    \circled{1} timestamps at job submission, start, and completion; 
    \circled{2} resource requests; 
    \circled{3} assigned node IDs; 
    \circled{4} \i{list of software dependencies} (indicative of the ML model/framework used).
    \circled{5} \i{user name and job name} (anonymized through hashing, yet still useful for inferring relationships between jobs, such as resubmissions); 
    \item \b{Node Status (aggregated for every 5/10/30 minutes)}: 
    \circled{1} node ID; 
    \circled{2} average/maximum CPU and host memory utilization rates; 
    \circled{3} average/maximum GPU and memory utilization rates; 
    \circled{4} \i{inbound/outbound traffic in bytes}; 
    \circled{5} \i{networked file storage IO in bytes}.
    \item \b{Node Specifications}: 
    \circled{1} Node ID; 
    \circled{2} GPU model; 
    \circled{3} number of GPUs; 
    \circled{4} total memory and other hardware specifications.
\end{ul}

The \i{italicized items} are unique data points that are not available in previous traces.

\t{Software Stack.} Our software stack includes the source code and configuration scripts for the cluster workload manager, scheduler, and monitoring tools. This includes:
\begin{ul}
    \item \sys-cli and \sys-service (\S\ref{section:design:layer2}) code and makefiles.
    \item Scheduler implementation (\S\ref{section:design:layer3}) as a Slurm plugin.
    \item Compute node (\S\ref{section:design:layer4}) configs (e.g., SR-IOV, GlusterFS).
\end{ul}

\t{Executor Backend Extension in Kubernetes Community.} 
SING's execution layer is designed to be extensible, enabling support for new cluster managers, with Slurm as the default. Recently, developers from DaoCloud~\cite{daocloud}, a leading Kubernetes contributor and software vendor, submitted a merge request to SING’s open-source repository~\cite{tacc-k8s}. 
This update adds support for job execution via Kubernetes and ports the scheduler to use the Kubernetes API. The development was presented in a report at KubeCon 2024~\cite{kubecon}, a Linux Foundation community event, as a step toward integrating cloud-native infrastructure with AI.

\section{Limitations}
\label{section:limitations}

\sys's design has the following known limitations:

\t{Resource Isolation}. Slurm isolates resources assigned to different jobs on the same host using Linux $\texttt{cgroups}$~\cite{slurm-isolation}. We currently depend on Slurm's built-in mechanisms for isolating resources assigned to various jobs. For container-format jobs, the container runtime is also invoked by Slurm within this isolation mechanism. The performance and security implications—specifically, whether user resources are entirely separated—require further research.

\t{Connection Between \sys-cli and \sys-service}. Currently, \sys-cli establishes a connection to \sys-service via SSH to execute remote commands. The user's SSH public key is added to the host running \sys-service, associated with a new Linux user created during registration. While this connection is secure, there remains a potential vulnerability where a malicious user could exploit it to execute unauthorized commands on the \sys-service host. Although we have restricted user permissions on the host system to allow only specific operations (such as invoking \sys-service), a more secure approach would be to expose \sys-service functionalities through a secure API with authentications.

\t{Port Forwarding for User Jobs}. To enable users to access interactive services (e.g., TensorBoard) running within their jobs, we employ port forwarding on the cluster gateway~(\S\ref{section:design:layer2}). This approach exposes a port on the \sys-service host, which may allow unauthorized access if the service lacks authentication. A more secure alternative is utilizing SSH tunneling~\cite{sshtunneling}. This method creates a secure tunnel between the job host and the user's local machine via the gateway, restricting access exclusively to the user.

\t{Cluster Failure}. \sys does not automatically recover from cluster failures, such as node failures. When a node fails, it and any associated user jobs become disconnected from the cluster, and require manual recovery by the cluster operator. Consequently, the job loses progress (up to the last checkpoint) and must be resubmitted.

\vspace{-6pt}
\section{Related Work and Discussion}
\label{section:discussion}

We discuss the distributed frameworks and schedulers for ML workloads. \sys can extend with these capabilities, leveraging advancements in storage, networking, and hardware for efficient, flexible cluster management.

\t{Distributed Computing Frameworks.}  
Distributed computing frameworks such as Ray~\cite{ray} and Spark~\cite{spark} perform computations at a granular level, allowing them to parallelize large jobs for efficient distributed execution. For example, Ray employs an actor and task-based model for more precise task management and asynchronous execution. These frameworks are integrated directly into user code, and submitted as part of the user's job to cluster managers.
We note that these frameworks can optionally operate in cluster mode~\cite{ray-cluster,spark-cluster}, functioning similarly to cluster managers by managing and allocating cluster resources to jobs. However, when running in cluster mode, submitted jobs need to use their computation APIs. This approach differs from traditional cluster managers like Slurm, Kubernetes, or \sys, which allow users to submit code using any programming framework.

\t{Cluster Schedulers.}  
Cluster schedulers for ML, including the state-of-the-art Pollux~\cite{pollux} and Sia~\cite{sia}, focus on job scheduling to improve cluster-wide performance such as job completion time. These schedulers can be integrated into cluster managers like Slurm, Kubernetes, and \sys to manage the entire ML job lifecycle. However, as discussed in our scheduler design in \S\ref{section:design:layer3}, \sys generally processes requests in a first-come, first-served (FCFS) manner to ensure equal access for any individual users.

\t{ML System Optimizations.} 
\sys is extensible to leverage advancements in system optimizations for ML, such as faster distributed storage~\cite{bufferedio} and more efficient networking~\cite{srnic, liteflow, pias}. These optimizations typically operate at layers orthogonal to \sys and can enhance ML job performance transparently. 
Additionally, \sys's scheduling policies can be extended to accommodate user-defined requirements or job profiling information. This will enable \sys to more effectively optimize job placement across heterogeneous nodes, including those with specialized hardware (e.g., TPU~\cite{tpu} and FPGA~\cite{codenet, flash}), as well as in geographically distributed environments (\S\ref{section:design:layer4}).

\t{Using Slurm as Executor Backend.}
Slurm, originally designed for MPI applications in scientific computing, is valued for its simplicity and performance. While the MPI model has gained less focus with the rise of big data and internet services, ML training jobs share system-level similarities with MPI, leveraging advanced MPI techniques~\cite{nccl, horovod}. As described in this paper, by adding ML workflow features to Slurm, we leverage Slurm's architecture and extend it to manage shared ML clusters. Kubernetes can also serve as the executor backend (\S\ref{section:opensource}).

\vspace{-6pt}
\section{Conclusion}

This paper addresses the pressing challenges faced by universities worldwide in effectively managing shared GPU clusters for ML research. 
Our experience with the design and operation of \sys, a campus ML cluster manager, offers an operation-efficient solution, reducing the deployment and maintenance burden on operations teams while achieving high resource utilization and user satisfaction.

\begin{acks}
We thank our shepherd, Malte Schwarzkopf, and the anonymous reviewers from ASPLOS 2025, as well as reviewers from our previous submissions, for their valuable feedback.
We also thank the Kubernetes community contributors, Peter Pan~(@panpan0000) and Xiao Zhang~(@wawa0210) from DaoCloud~\cite{daocloud}, for their insightful discussions and valuable contributions to the implementation of \sys.

\sys was developed, deployed, and operated on the TACC cluster~\cite{tacc} at The Hong Kong University of Science and Technology. The statistics and analysis reported in this paper are also based on SING's deployment and operation on TACC. An arXiv paper~\cite{tacc-arxiv} detailing the TACC architecture, which includes a bottom-up approach to efficient AI computing with networking, frameworks, and algorithms, forms the foundation and inspiration of this work.

This work is supported in part by the Hong Kong RGC TRS T41-603/20R, GRF 16213621, NSFC 62062005, 62402407. Kai Chen is the corresponding author.
  
\end{acks}

    \bibliographystyle{ACM-Reference-Format}
    \balance
    \bibliography{references}
\end{document}